\documentclass[aps,showpacs, superscriptaddress,preprint,preprintnumbers,amsmath,amssymb]{revtex4}
\usepackage{graphicx}
\usepackage{dcolumn}
\usepackage{bm}
\begin{document}
\renewcommand{\thefootnote}{\fnsymbol{footnote}}
\title{Bond-breaking Induced Lifshitz Transition in Robust Dirac Semimetal $\mathbf{VAl_3}$}

\author{Yiyuan Liu$^\star$}
\affiliation{International Center for Quantum Materials, School of Physics, Peking University, China}

\author{Yu-Fei Liu} 
\affiliation{International Center for Quantum Materials, School of Physics, Peking University, China}

\author{Xin Gui}
\affiliation{Department of Chemistry, Louisiana State University, Baton Rouge, LA 70803, USA}

\author{Cheng Xiang}
\affiliation{International Center for Quantum Materials, School of Physics, Peking University, China}

\author{Hui-bin Zhou} 
\affiliation{International Center for Quantum Materials, School of Physics, Peking University, China}

\author{Chuang-Han Hsu}
\affiliation{Departments of Physics, National University of Singapore, Singapore 117542}
\affiliation{Centre for Advanced 2D Materials and Graphene Research Centre, National University of Singapore, Singapore 117546}

\author{Hsin Lin}
\affiliation{Institute of Physics, Academia Sinica, Taipei 11529, Taiwan}

\author{Tay-Rong Chang}
\affiliation{Department of Physics, National Cheng Kung University, Tainan 701, Taiwan}
\affiliation{Center for Quantum Frontiers of Research \& Technology (QFort), Tainan 701, Taiwan}
\affiliation{Physics Division, National Center for Theoretical Sciences, Hsinchu, Taiwan}

\author{Weiwei Xie}
\affiliation{Department of Chemistry, Louisiana State University, Baton Rouge, LA 70803, USA}

\author{Shuang Jia}
\email{gwljiashuang@pku.edu.cn}
\affiliation{International Center for Quantum Materials, School of Physics, Peking University, China}
\affiliation{Collaborative Innovation Center of Quantum Matter, Beijing 100871, China}
\affiliation{CAS Center for Excellence in Topological Quantum Computation, University of Chinese Academy of Science, Beijing 100190, China}
\affiliation{Beijing Academy of Quantum Information Sciences, West Building 3, No. 10 Xibeiwang East Road, Haidian District, Beijing 100193,China}

\clearpage

\begin{abstract}
Topological electrons in semimetals are usually vulnerable to chemical doping and environment change, which restricts their potential application in future electronic devices. In this paper we report that the type-\uppercase\expandafter{\romannumeral2} Dirac semimetal $\mathbf{VAl_3}$ hosts exceptional, robust topological electrons which can tolerate extreme change of chemical composition. The Dirac electrons remain intact even after a substantial part of V atoms have been replaced in the $\mathbf{V_{1-x}Ti_xAl_3}$ solid solutions. This Dirac semimetal state ends at $x=0.35$ where a Lifshitz transition to $p$-type trivial metal occurs. The V-Al bond is completely broken in this transition as long as the bonding orbitals are fully depopulated by the holes donated from Ti substitution. In other words, the Dirac electrons in $\mathbf{VAl_3}$ are protected by the V-Al bond whose molecular orbital is their bonding gravity center. Our understanding on the interrelations among electron count, chemical bond and electronic properties in topological semimetals suggests a rational approach to search robust, chemical-bond-protected topological materials.

\end{abstract}

\maketitle
\section{Introduction}
Topological semimetals (TSMs) host relativistic electrons near band-crossing points in their electronic structures \cite{X.-L.Qi2011,M.Z.Hasan2011,RevModPhys.90.015001}. These electrons' low-energy excitation obeys the representations of Dirac equation in particle physics and therefore they are dubbed as Weyl and Dirac fermions. The topologically protected electrons are highly mobile because their topological state is robustly against small local perturbations. In contrast, the chemical potential in TSMs is very sensitive to small change of composition and external environment. Thus tiny defect may depopulate electrons away from the topological band.
This vulnerability limits the TSMs' potential application in future electronic devices.

Here we report robust topological electrons in Dirac semimetal (DSM) $\mathrm{VAl_3}$ whose electronic structure can tolerate an extreme chemical composition change. We find that the $\mathrm{V_{1-x}Ti_xAl_3}$ solid solutions feature standard transport behaviors of DSM even after a substantial part of V atoms ($35\%$) have been replaced.
Titanium substitution induces a Lifshitz transition from $n$-type DSM to $p$-type trivial metal as long as the V-Al bond is completely broken. This Lifshitz transition is controlled by the covalent V-Al bond which shields the Dirac electrons robustly.

Vanadium and titanium trialuminide crystallize in a same structure which is built from the arrangement of the Al$_{12}$ cuboctehedra containing V$/$Ti atoms (Inset in Fig.~4).
They belong to a group of polar intermetallics in which strong metallic bonding occurs within the covalent partial structure \cite{miller2011quantitative,guloy2006polar,vajenine1998magic}.
Previous studies on their band structure and molecular orbitals have clarified that this crystal structure is stabilized by a "magic number" of 14 electrons per transition metal when a pseudo gap occurs in the density of states (DOS) \cite{jahnatek2005interatomic,colinet2001phase,boulechfar2010fp,krajci2002covalent,chen2010chemical}.
Recent theoretical work demonstrated that there exist a pair of tilted-over Dirac cones in the pseudo gap in the energy-momentum space slightly above the Fermi level (E$_f$) of VAl$_{3}$ \cite{chang2017type,chen2018bulk}. The tilted-over Dirac cones host type-\uppercase\expandafter{\romannumeral2} Lorentz-symmetry-breaking Dirac fermions \cite{soluyanov2015type,wang2016mote}, which can give rise to many exotic physical properties, such as direction-dependent chiral anomaly and Klein tunneling in momentum space \cite{udagawa2016field,o2016magnetic}. Very recently the Fermi surface and the topological planar Hall effect (PHE) in $\mathrm{VAl_3}$ were illustrated in experiment \cite{chen2018bulk,singha2018planar}. So far, no relevant study has considered the interrelations among electron count, chemical bond and topological properties in $\mathrm{VAl_3}$.

In this paper we focus on the crystal structure and electronic properties of the isostructural solid solutions of $\mathrm{V_{1-x}Ti_xAl_3}$ in which the electron count changes from $14$ to $13$. We observe a V-Al bond-breaking-induced lattice distortion. Concomitantly, the Hall signal changes from $n$-type to $p$-type at $x=0.35$. Further investigation reveals that the topological properties such as PHE in $\mathrm{VAl_3}$ remain intact after substantial Ti substitution, until a Lifshitz transition occurs. By performing electronic structure calculation with the emphasis on the crystal orbital Hamilton population (COHP), we reveal that the maximizing bonds, in particular the interplanar V-Al bond, are responsible for the structure distortion. The bond formation attempts to lower the DOS at E$_f$ and protects the type-\uppercase\expandafter{\romannumeral2} Dirac fermions when $x$ is less than $0.35$. The V-Al antibonding orbital builds up the Dirac electron band which is fully depopulated as long as the Lifshitz transition occurs. The relationship among the structure, electron count and electronic properties in various TSMs \cite{gibson2014quasi,lin2010half,muchler2012topological} has been considered, but the influence of the presence or absence of a chemical bond on topological electrons is less pondered \cite{seibel2015gold}. Our finding highlights a chemical bonding gravity center \cite{hoff1988} of topological electrons in particular TSMs. Understanding this chemical protection mechanism in various TSMs will help to develop more robust electronic devices which may have potential application in the future.

\section{Result}

\begin{figure}
	\centering
	\includegraphics[width=1\linewidth]{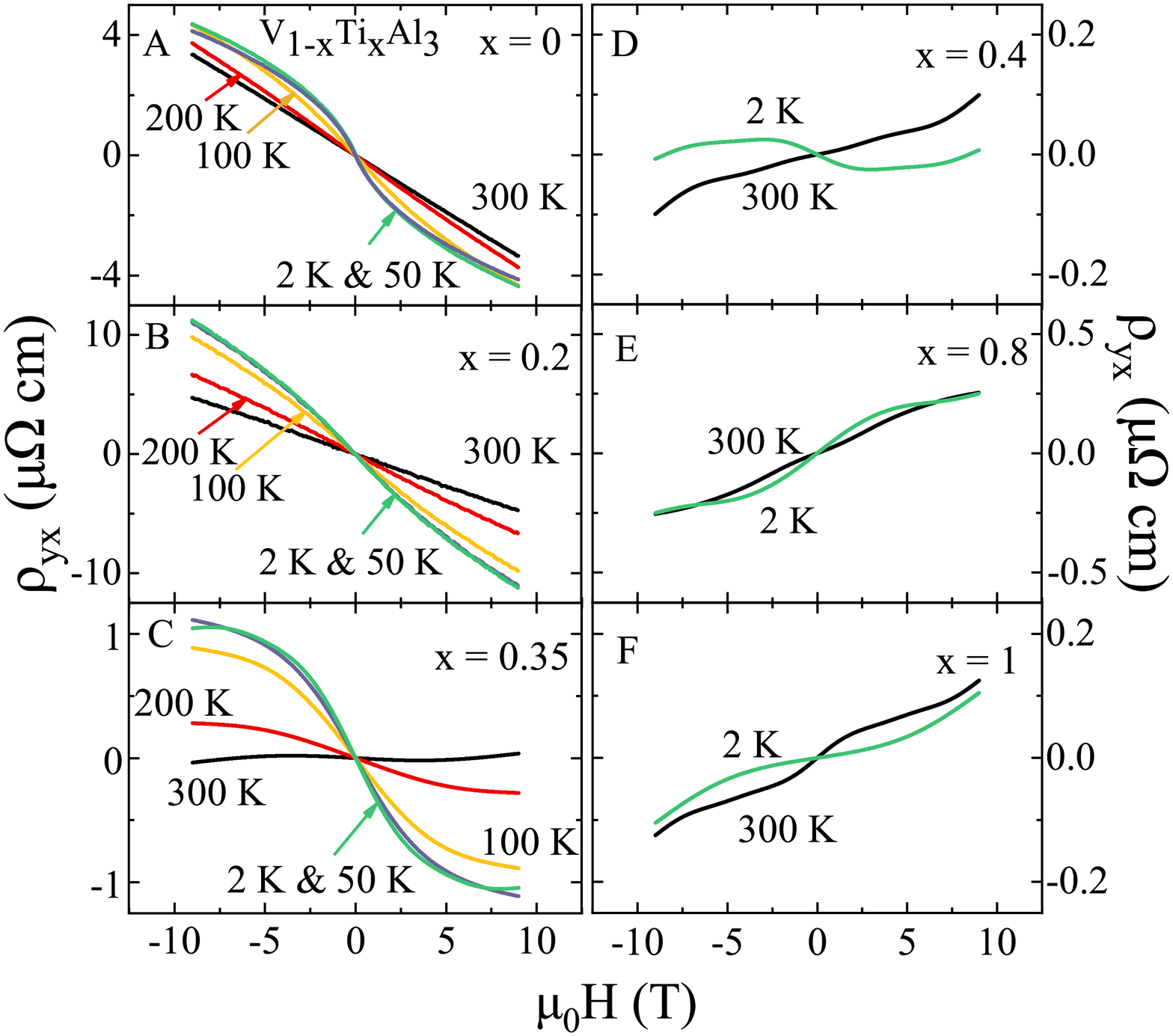}
	\caption{Field-dependent Hall resistivity ($\rho_{yx}$) at different temperatures for representative samples in $\mathrm{V_{1-x}Ti_xAl_3}$. (A), (B) and (C): $x=0,0.2$ and $0.35$ at 2 K, 50 K, 100 K, 200 K, and 300 K, respectively. The data at $2$ K (green) and $50$ K (purple) are nearly identical. (D), (E) and (F): $x=0.4,0.8$ and $1$ at 2 K and 300 K, respectively. }
	\label{fig:1}
\end{figure}

We firstly demonstrate the electronic properties of $\mathrm{V_{1-x}Ti_xAl_3}$. Figure \ref{fig:1} shows the Hall resistivity ($\rho_{yx}$) with respect to the external magnetic field ($H$) at different temperatures for representative samples. $\mathrm{VAl_3}$ is a $n$-type semimetal whose $\rho_{yx}$ is large and negatively responsive to field \cite{singha2018planar}. $\rho_{yx}$ is not linearly dependent on $H$ below $50$ K, indicating two types of carriers coexist. If we adopt a rigid-band approximation, a small Ti substitution ($3.5\%$) is expected to compensate the destitute conduction electrons in $\mathrm{VAl_3}$ and leads to a $n$-$p$ transition. Surprisingly, the profile of $\rho_{yx}$ remains almost intact after $35\%$ of V atoms are replaced. We find a robust semimetal state in $\mathrm{VAl_3}$ which can tolerant large chemical composition change. 

This semimetal state ends at $x=0.35$. When $x=0.4$, $\rho_{yx}$ is very small because the lowly mobile electrons and holes are compensated. Yet we can discern that the $\rho_{yx}$ is positively dependent on $H$ at $300$ K but negatively dependent on $H$ at $2$ K, indicating a $n$-$p$ transition onsite. When $x$ is more than $0.4$, the alloy becomes a $p$-type metal showing a small and linear, temperature-independent $\rho_{yx}$.

\begin{figure}
	\centering
	\includegraphics[width=.8\linewidth]{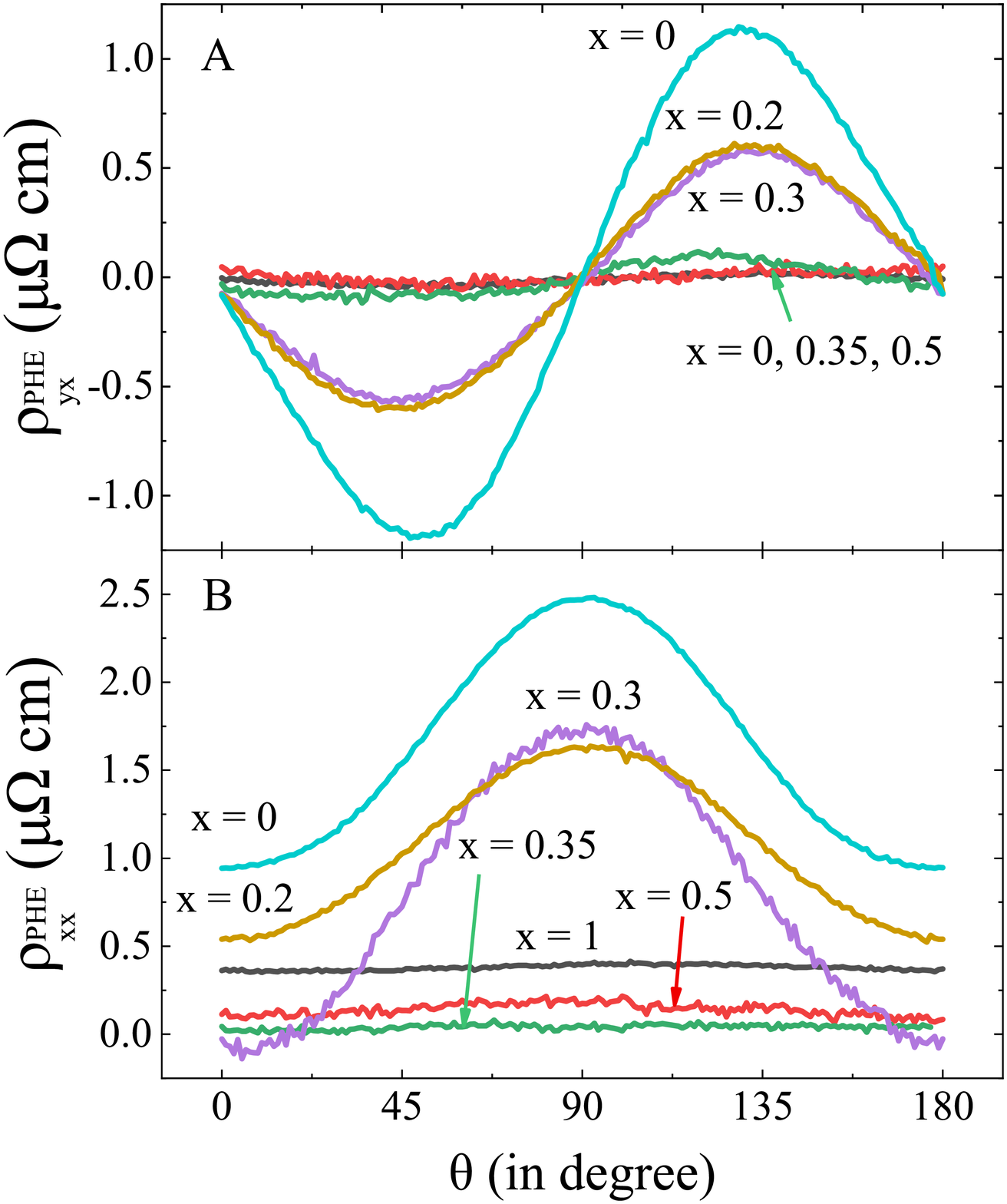}
	\caption{(A) Planar Hall resistivity ($\rho_{yx}^\mathrm{PHE}$) and (B) anisotropic magneto-resistivity ($\rho_{xx}^\mathrm{PHE}$) with respect to the angle $\rm{\theta}$ in a magnetic field of 5 T at 2 K for representative samples in $\mathrm{V_{1-x}Ti_xAl_3}$.}
	\label{fig:2}
\end{figure}

\begin{figure}
	\centering
	\includegraphics[width=0.8\linewidth]{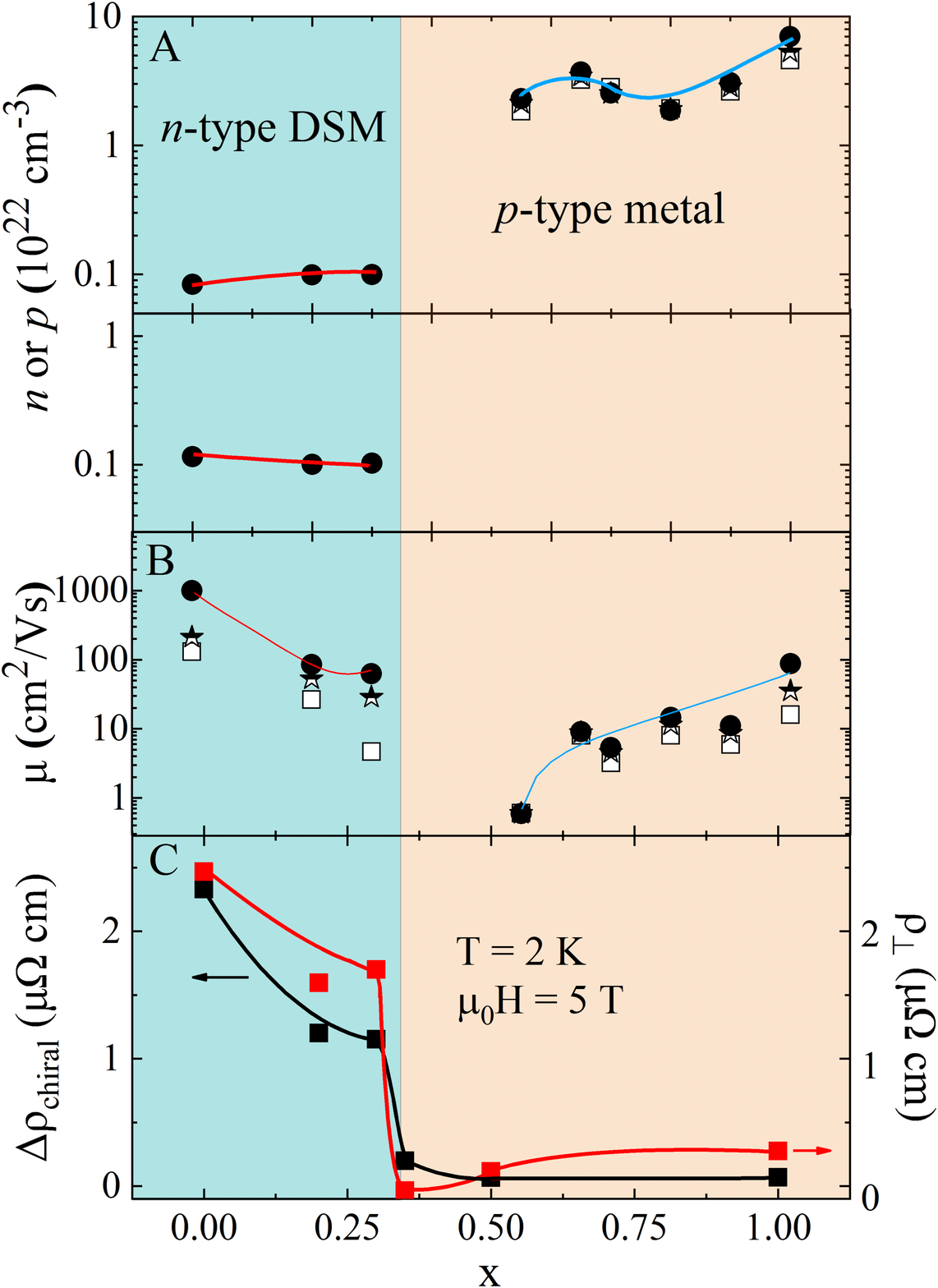}
	\caption{(A) Carrier density, (B) mobility for $\mathrm{V_{1-x}Ti_xAl_3}$. The solid circles, semisolid stars and hollow squares represent the data at $2$ K, $150$ K and $300$ K, respectively. (C) $\Delta\rho_\mathrm{chiral}$ and $\rho_{\perp}$ in a magnetic field of 5 T at 2 K.}
	\label{fig:3}
\end{figure}

We then investigate the PHE in $\mathrm{V_{1-x}Ti_xAl_3}$ which is known as an evidence of existing Dirac fermions \cite{singha2018planar,burkov2017giant}. The planar Hall resistivity ($\rho_{yx}^\mathrm{PHE}$) and anisotropic magneto-resistivity ($\rho_{xx}^\mathrm{PHE}$) in DSMs can be described as the following formula \cite{burkov2017giant,nandy2017chiral},

\begin{center}
	\begin{equation}
	\rho_{yx}^\mathrm{PHE}=-\Delta\rho_\mathrm{chiral}sin\theta cos\theta
	\label{eq1}
	\end{equation}
\end{center}
\begin{center}
	\begin{equation}
	\rho_{xx}^\mathrm{PHE}=\rho_{\perp}-\Delta\rho_\mathrm{chiral}cos2\theta
	\label{eq2}
	\end{equation}
\end{center}

where $\rho_{\perp}$ and  $\rho_{\parallel}$ are the resistivity for current perpendicular and parallel to the magnetic field direction, respectively; $\Delta\rho_\mathrm{chiral}=\rho_{\perp}-\rho_{\parallel}$ gives the anisotropy in resistivity induced by chiral anomaly. Previous study showed large PHE in $\mathrm{VAl_3}$ \cite{singha2018planar} and here we focus on its change when Ti atoms are substituted. As shown in Fig.~\ref{fig:2}, the $\rho_{yx}^\mathrm{PHE}$ and $\rho_{xx}^\mathrm{PHE}$ show explicit periodic dependence on $\theta$ which can be well-fitted by using Eqs.~(\ref{eq1}) and (\ref{eq2}) when $x$ is less than $0.35$. Remarkably the angular dependent $\rho_{yx}^\mathrm{PHE}$ and $\rho_{xx}^\mathrm{PHE}$ suddenly drop to almost zero at $x=0.35$, which is coincident with the $n$-$p$ transition. We extract $\Delta\rho_\mathrm{chiral}$ and $\rho_{\perp}$ for each $x$ as shown in Fig.~\ref{fig:3}.

The carrier concentration ($n$ and $p$ for electron and hole, respectively) and mobility ($\mu$) for $\mathrm{V_{1-x}Ti_xAl_3}$ are obtained by fitting the field-dependent resistivity and $\rho_{yx}$ at different temperatures. As shown in Fig.~\ref{fig:3}, the carrier concentration in the V-rich semimetal region is nearly invariant as $x$ increases from $0$ to $0.3$ while the mobility is relatively large during the process.
As comparison, because the carrier density and electron scattering are governed by increasing charge defects 
\cite{ziman2001electrons} in topological trivial alloys in the Ti-rich region, the hole mobility drops significantly when $x$ changes from $1$ to $0.55$.
Concomitant with the $n$-$p$ transition, the $\Delta\rho_\mathrm{chiral}$ and $\rho_{\perp}$ dramatically drop to almost zero when $x$ is more that $0.35$. Previous studies suggested that the PHE and angular dependent resistivity are determined by the Berry curvature in non-magnetic TSMs \cite{nandy2017chiral, li2018giant}.
Our observation reveals that the $n$-$p$ transition at $x=0.35$ is a topological transition from a DSM to a trivial metal.

\begin{figure}
	\centering
	\includegraphics[width=.8\linewidth]{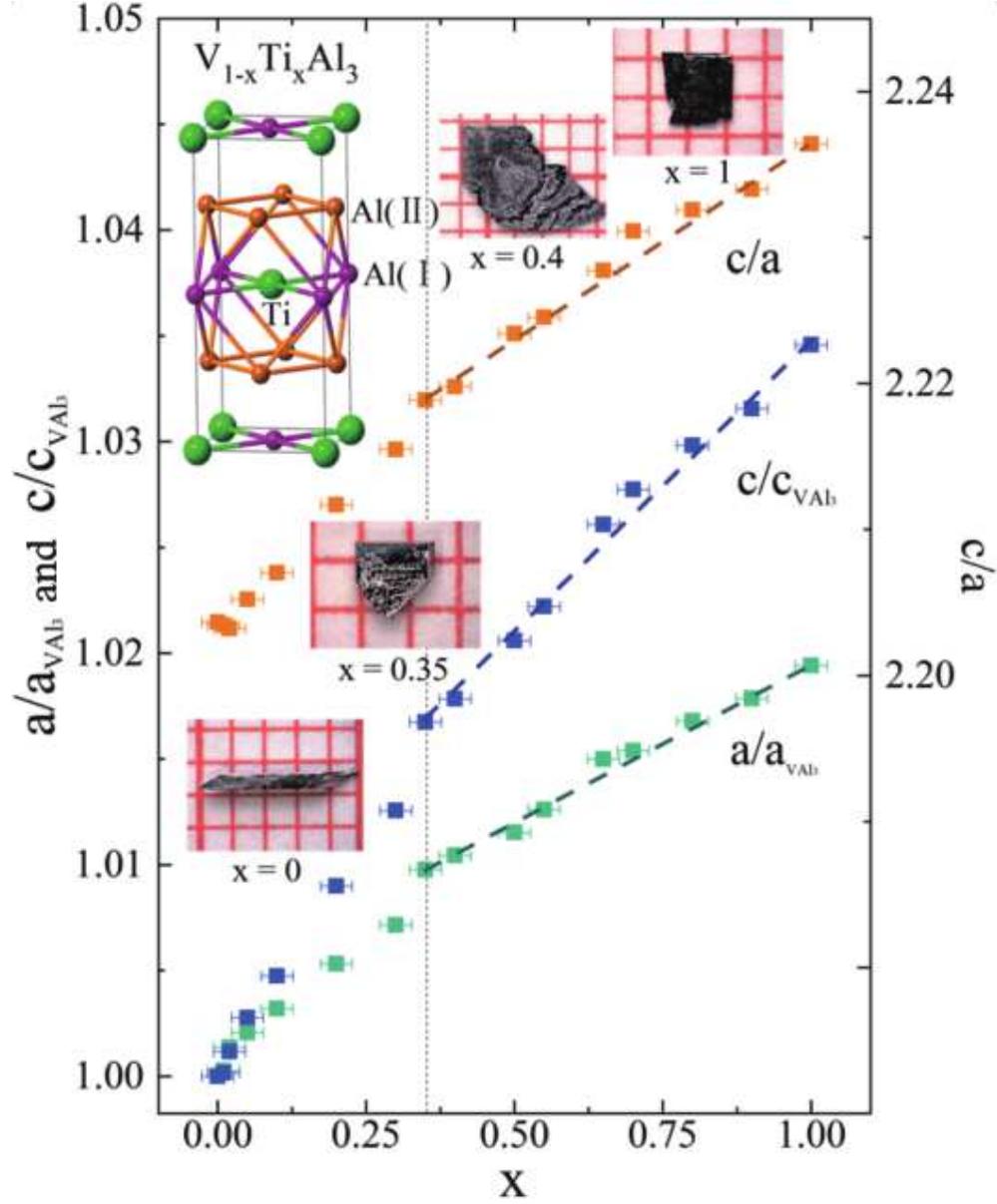}
	\caption{Lattice parameters ($a$ and $c$) for $\mathrm{V_{1-x}Ti_xAl_3}$. The error bar of $x$ is estimated as $\pm2\%$ while the error bar of $a$ and $c$ ($<3\times10^{-3}\AA$) is smaller than the square data point . The straight dashed lines are guided by eye. Inset in the top left corner shows the unit cell of $\mathrm{TAl_3}$. The insets from left to right show photos of the single crystals for $x=0, 0.35, 0.4$ and $1$, respectively.}
	\label{fig:4}
\end{figure}

The unusual evolution of electronic properties in the isostructural solutions motivates us to exam their crystal structure change. Before doing any characterization, we can easily distinguish the rod-like single crystals of $\mathrm{VAl_3}$ from the plate-like $\mathrm{TiAl_3}$ at the first sight (Insets in Fig.~\ref{fig:4}). The square plate of $\mathrm{TiAl_3}$ shows glossy $(001)$ facets and such crystal shape is commonly observed in the compounds composed by slender tetragonal unit cells. As comparison, the rod-like crystals of $\mathrm{VAl_3}$ have four glossy $(110)$ rectangle facets. The insets of Fig.~\ref{fig:4} show that the crystal shape evolves from a flat plate for $x=1$, to a three dimensional chunk for $x=0.35$, and then to a long rod for $x=0$. Moreover, the lattice parameters $a$, $c$ and the ratio of $c/a$ shrink linearly with respect to $x$ when $x$ is more than $0.35$, but they change accelerately when $x$ is less than $0.35$ (See Fig.~\ref{fig:4} and SI Appendix, Fig. S4). This curious nonlinear change is less than $2\%$ but it cannot be ascribed to a chemical pressure effect coming from the different size of V and Ti atoms. Contrary to our expectation, the crystal structure is seriously distorted for $x<0.35$ whereas the electrical transport properties remain intact in this region. In the following part, we will prove that a chemical bond breaking plays a crucial role in the evolution of electronic properties and crystal structure.

\section{Discussion}

The crystal structures and the pseudo gap in various transition metal trialuminide ($\mathrm{TAl_3}$) have attracted much attention in chemistry \cite{condron2003new,freeman1990phase,xu1989band,hong1990crystal,xu1990phase}. Yannello $\it{et.al.}$ suggested a bonding picture in which each T atom is connected by 4 T-T sigma bonds through the T-Al supporting (See SI Appendix, Fig. S1)  \cite{kilduff2015defusing,yannello2015generality,yannello2014isolobal}. Therefore the electron count needs $18-4=14e^-$ to achieve a closed-shell configuration. Ironically, $\mathrm{TiAl_3}$, as the prototype of this crystal structure, fails to fulfill the criteria because it has only $13e^-$ per formula. We expect a transition from semimetal to metal in $\mathrm{V_{1-x}Ti_xAl_3}$ , but the relationship among electron count, structural distortion and electronic properties needs further elaboration in theory.

\begin{figure*}
	\begin{center}
		\includegraphics[width=1\linewidth]{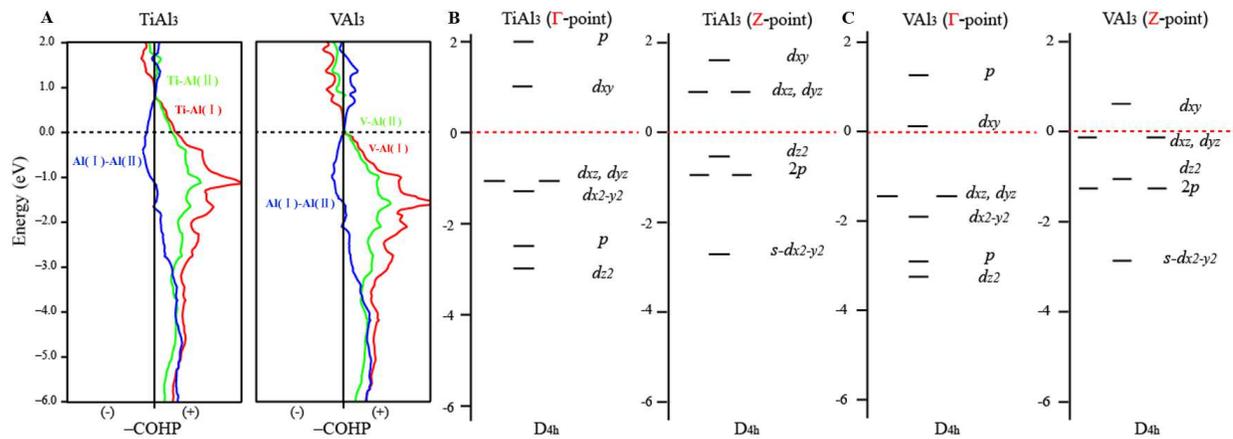}
		\caption{(A) Crystal orbital Hamilton population (COHP) for $\mathrm{TiAl_3}$ and $\mathrm{VAl_3}$. (B) and (C) Molecular energy level diagrams at $\Gamma$ and $Z$ points for $\mathrm{TiAl_3}$ and $\mathrm{VAl_3}$, respectively.}
		\label{fig:5}
	\end{center}
\end{figure*}

To shed a light on the relation between electron count and chemical bond, we compare the electronic structures between $\mathrm{TiAl_3}$ and $\mathrm{VAl_3}$. As illustrated in Fig. \ref{fig:5}, the lone pair electrons from V atoms make the chemical bonding state and the valence band in momentum space just fully filled at the same time. This indicates that $14$ valence electrons ($5+3\times$3 for $\mathrm{VAl_3}$) optimize the bonding in this structure. By reducing one electron, $\mathrm{VAl_3}$ degrades from fully saturated bonding state to partially saturated state for $\mathrm{TiAl_3}$. To clarify the orbital contributions, a schematic picture based on the molecular perspective of $\mathrm{TiAl_3}$ and $\mathrm{VAl_3}$ is also presented in Fig. \ref{fig:5}. The $d$ orbitals on transition metals (Ti or V) split into a$_{1g}$(d$_{z^2}$), b$_{1g}$(d$_{x^2-y^2}$), e$_g$(d$_{xz}$, d$_{yz}$) and b$_{2g}$(d$_{xy}$) with the E$_f$ locating between e$_g$ and b$_{2g}$. Interestingly, the orbital splitting between e$_g$(d$_{xz}$, d$_{yz}$) and b$_{2g}$(d$_{xy}$) significantly reduces as long as the band inversion occurs between a$_{1g}$(d$_{z^2}$) and b$_{1g}$(d$_{x^2-y^2}$) from $\Gamma$-point to $Z$-point in the brillouin zone (BZ) (Fig. \ref{fig:7} ). Given the above, there exists a covalent V-Al bond in $\mathrm{VAl_3}$ which generates a pseudo-gap between e$_g$ and b$_{2g}$ orbitals and that gap is minimized at $Z$-point in the BZ. The related molecular orbitals in $\mathrm{VAl_3}$ and $\mathrm{TiAl_3}$ see SI appendix, Fig. S3.

\begin{figure}
	\centering
	\includegraphics[width=.8\linewidth]{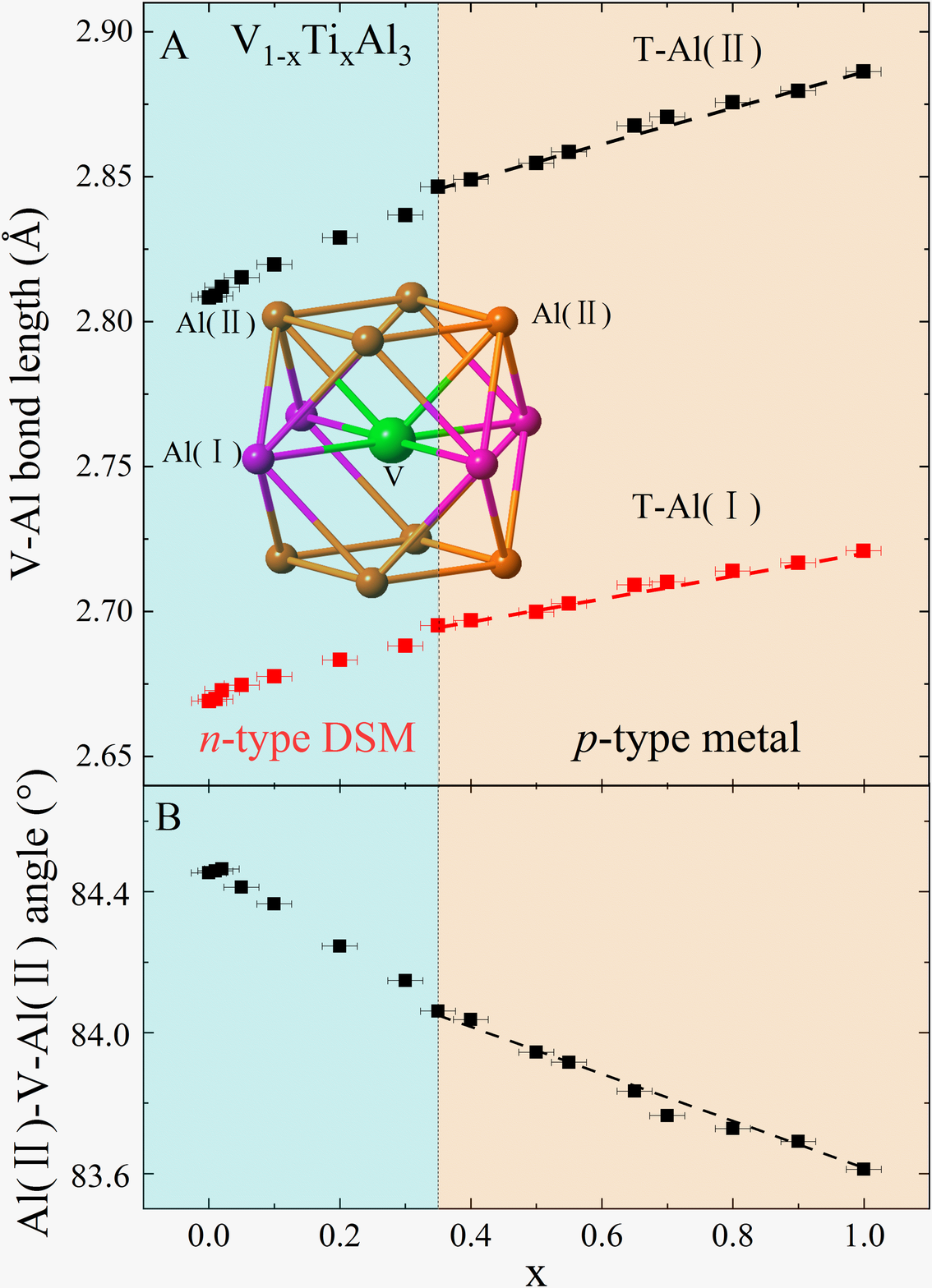}
	\caption{(A) T-Al(\uppercase\expandafter{\romannumeral1}) and T-Al(\uppercase\expandafter{\romannumeral2}) bond length and (B) Al(\uppercase\expandafter{\romannumeral2})-T-Al(\uppercase\expandafter{\romannumeral2}) bond angle for $\mathrm{V_{1-x}Ti_xAl_3}$. Inset shows the V-centered Al$_{12}$ cuboctahedra. The straight dashed lines are guided by eye.}
	\label{fig:6}
\end{figure}

\begin{figure}
	\centering
	\includegraphics[width=1\linewidth]{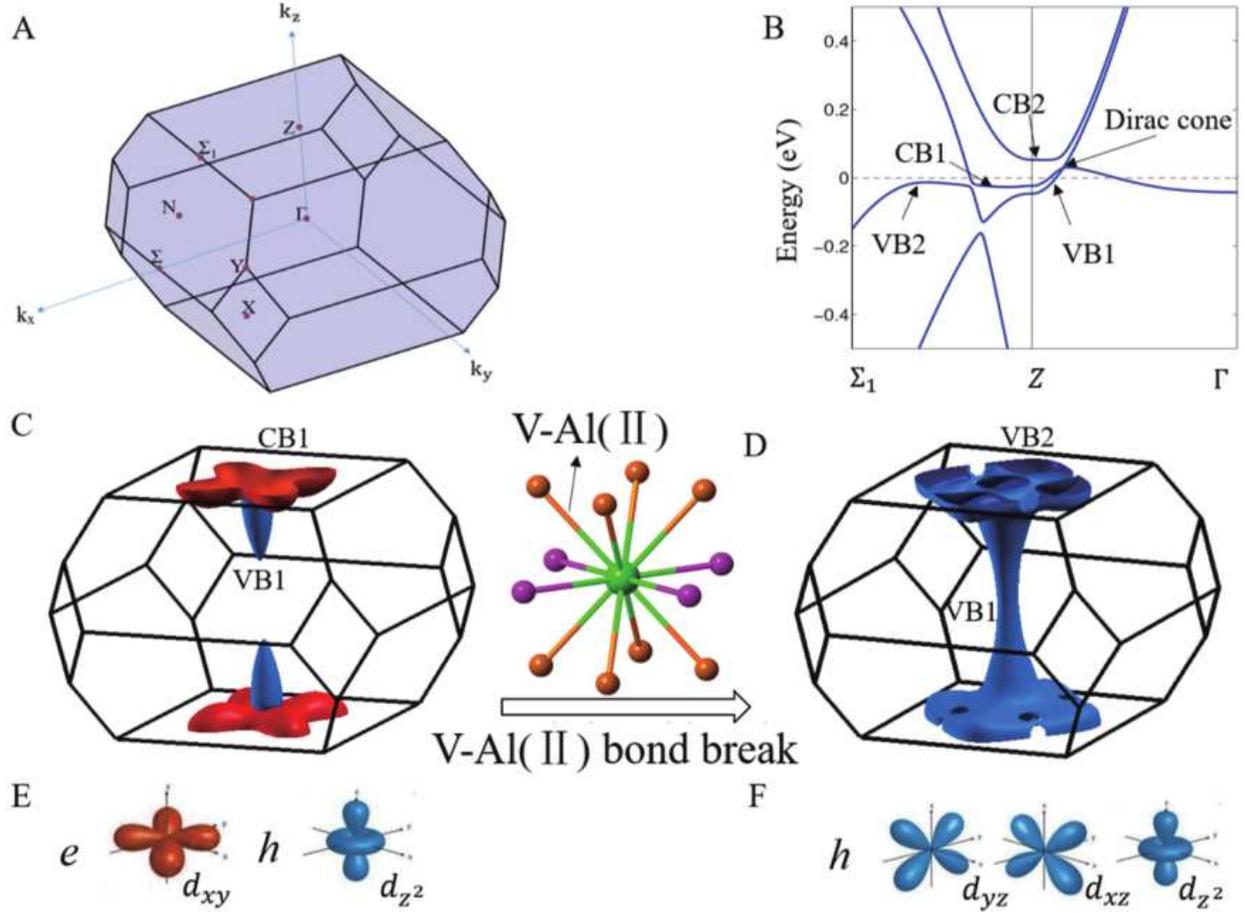}
	\caption{(A) Brillouin zone of $\mathrm{VAl_3}$. (B)Band structure in the vicinity of the type-\uppercase\expandafter{\romannumeral2} Dirac node in $\mathrm{VAl_3}$. (C) Electron (CB1, red) and hole (VB1, blue) pockets of $\mathrm{VAl_3}$. (D) Hole pockets (VB1 and VB2, blue) in $\mathrm{V_{1-x}Ti_xAl_3}$ for $x>0.35$ as long as the V-Al(\uppercase\expandafter{\romannumeral2}) bond breaks. (E) $d_{xy}$ (red) and $d_{z^2}$ (blue) orbitals compose CB1 and VB1, respectively. (F) Degenerated $d_{yz}$/$d_{xz}$ and $d_{z^2}$ orbitals compose VB2 and VB1, respectively.}
	\label{fig:7}
\end{figure}

Figure \ref{fig:6} shows the change of inter-planar T-Al(\uppercase\expandafter{\romannumeral2}) and inner-planar T-Al(\uppercase\expandafter{\romannumeral1}) bond length and Al(\uppercase\expandafter{\romannumeral2})-T-Al(\uppercase\expandafter{\romannumeral2}) bond angle for the whole series. Although both of the T-Al(\uppercase\expandafter{\romannumeral1}) and T-Al(\uppercase\expandafter{\romannumeral2}) bonds elongate when $x$ increases, T-Al(\uppercase\expandafter{\romannumeral2}) bond changes much more than T-Al(\uppercase\expandafter{\romannumeral1}) bond for $x<0.35$, and this change makes the Al(\uppercase\expandafter{\romannumeral2})-T-Al(\uppercase\expandafter{\romannumeral2}) bond angle much wider for $x<0.35$.
Combining the COHP calculation in Fig. \ref{fig:5}, we derive that the T-Al(\uppercase\expandafter{\romannumeral2}) bond plays a crucial role because it has more adjustable length and bonding orientation compared with the weak Al(\uppercase\expandafter{\romannumeral1})-Al(\uppercase\expandafter{\romannumeral2}) bond and rather stable V-Al(\uppercase\expandafter{\romannumeral1}) bond. The Ti-substitution successively weakens the V-Al(\uppercase\expandafter{\romannumeral2}) bond until $x=0.35$ beyond which the bond is fully broken. 

The change of the crystal shape in $\mathrm{V_{1-x}Ti_xAl_3}$ reflects the bond breaking as well. Remember all the crystals form in molten Al, the V-Al(\uppercase\expandafter{\romannumeral2}) dangling bonds should attract more atoms along the $c$ axis for $x<0.35$. Because the facets of crystals always intent to have fewer dangling chemical bonds, the strong V-Al(\uppercase\expandafter{\romannumeral2}) bond gives rise to the rod-like shape of $\mathrm{VAl_3}$ single crystals.

The bond breaking scheme above naturally explains the unusual change of the electronic properties in V$_{1-x}$Ti$_x$Al$_3$. Remember the Lifshitz transition point ($x=0.35$) divides the whole series into metal and semimetal regions in which the crystal structure and the electronic transport properties change in different ways (Fig. \ref{fig:3} and \ref{fig:4} ). We focus on the semimetal region in which the topological electrons remain intact in defiance of the large change of the unit cell. We check the band structure of $\mathrm{VAl_3}$ in which two conduction bands (CB1 and CB2) and two valence bands (VB1 and VB2) are near the E$_f$ (Fig. \ref{fig:7}). The conduction band CB1 crosses with the valence band VB1 along the $\Gamma-Z$ direction, forming the Dirac nodes near the $Z$ point which host the highly mobile Dirac electrons.
Another hole pocket of VB2 emerges along the $\Sigma_1-Z$ direction below $15$ meV of the E$_f$. Although these bands are formed by the hybridization of different orbitals, we can sleuth out their mainly component atomic orbitals by the symmetry in $k$ space. The cloverleaf shaped electron pocket of CB1 along the diagonal direction mainly stems from the b$_{2g}$(d$_{xy}$) orbital, while the hole pocket of VB2 along the straight direction mainly stems from the degenerated e$_g$(d$_{xz}$, d$_{yz}$) orbitals. On the other hand, the large dispersion along the k$_z$ direction in the VB1 top inherits the p$_z$ orbital of Al and a$_{1g}$(d$_{z^2}$) orbital of V. This rough estimation is consistent with our molecular energy level diagram on the $\Gamma$ and $Z$ points (Fig. \ref{fig:5} ). The $n$-$p$ transition occurs as long as the electrons fully depopulate the b$_{2g}$ orbital. This bond breaking corresponds to the Lifshitz transition in which the E$_f$ leaves the CB1 bottom and starts to touch the VB2 top. The V-Al bond is fully broken and the system degenerates to a trivial $p$-type metal as long as the VB2 starts to be populated when $x$ is more than $0.35$.

Hoffmann suggested that maximizing bond, acting as a reservoir to store electrons, always makes every effort to lower the DOS at the E$_f$ \cite{hoff1988}. When $x$ is less than $0.35$ in $\mathrm{V_{1-x}Ti_xAl_3}$, the bond formation demands extra electrons, which compensates the V-substitution and therefore the E$_f$ is pinned on the VB2 top. Vice versa, the Ti-substitution in $\mathrm{VAl_3}$ breaks the V-Al bond at first, and such bond breaking process acts as a buffer to retard the Lifshitz transition. In the scenario of the molecular orbitals, the covalent bond of V-Al(\uppercase\expandafter{\romannumeral2}) acts as the bonding gravity center \cite{hoff1988} of the topological electrons in $\mathrm{VAl_3}$. The type-\uppercase\expandafter{\romannumeral2} Dirac electrons are protected by the V-Al bond.

\section{Conclusion}
We observe robust Dirac electrons which are protected by the V-Al chemical bond in the $\mathrm{V_{1-x}Ti_xAl_3}$ solid solutions. When Ti atoms substitute V atoms, the V-Al bond acts as a shield for the Dirac electrons in $\mathrm{VAl_3}$. As long as the V-Al bond is completely broken, the $n$-type DSM degenerates to $p$-type trivial metal through a Lifshitz transition. We further infer that this kind of chemical bond protection is likely unique in type-\uppercase\expandafter{\romannumeral2} TSMs which host the tilted-over Dirac cones. In such band structure, the bond formation may partially populate the Dirac cone which acts as a dispersive part around the molecular-orbital-energy center. Finally, we suggest that the manipulation of chemical bond in topological materials bears more consideration for designing the functional quantum materials. Our future study in quantum materials will involve various chemical bonding models and electron counting rules.

\section*{Methods}
\subsection*{Synthesis}
Single crystals of $\mathrm{V_{1-x}Ti_xAl_3}$ were grown via a high-temperature-solution method with Al as the flux \cite{canfield1992growth}. The raw materials of V pieces ($99.99\%$), Ti powders ($99.99\%$) and Al ingot ($99.9\%$) were mixed together in a molar ratio of V : Ti : Al $= x: 1-x: 49$ and then placed in an alumina crucible which was sealed in a fused silica ampoule under partial argon atmosphere. The mixtures were heated up to $1323$ K for $10$ h to ensure the homogeneity. Then the crystal growth process involved the cooling from $1323$ K to $1023$ K over a period of $5$ days, followed by  decanting in a centrifuge. The typical size of $\mathrm{V_{1-x}Ti_xAl_3}$ series crystals varies from $2\times2\times0.1~\mathrm{mm}^3$ for plate-like to $0.5\times0.5\times5~\mathrm{mm}^3$ for rod-like.

\subsection*{Composition and Structure Determinations}
Powder XRD measurements were carried out in a Rigaku Mini-flux 600 diffractometer with Cu-Kα radiation. The diffraction data for $\mathrm{V_{1-x}Ti_xAl_3}$  series ($x=0,~0.05,~0.1,~0.2,~0.3,~0.35,~0.4,~0.5,~0.6,~0.7,~0.8,~0.9,~1$) were refined by Rietveld Rietica and the crystal parameters and atomic positions for $\mathrm{VAl_3}$ and $\mathrm{TiAl_3}$ were used as starting points. All refinements using Le Bail as the calculation method were quickly converged. The determination of the parameter $a$ and $c$ can completely describe the structure because all the atoms are located in high-symmetry positions [Al(\uppercase\expandafter{\romannumeral1}) $(0, 0, 0.5)$, Al(\uppercase\expandafter{\romannumeral2}) $(0, 0.5, 0.25)$ and V/Ti $(0, 0, 0)$]. The bond length and angle were obtained by geometric relations. We also applied single crystal XRD for the samples of $x=0,~0.2,~0.4,~0.6,~0.8,~1$ and the results see SI Appendix, Tables   S1 to S3.
Some literatures reported that the polycrystalline $\mathrm{TiAl_3}$ transforms to a complicated, superstructure at low temperature via a very sluggish and incomplete reaction \cite{loo1973diffusion,braun2001phase,karpets2003influence}.
We found that our single crystals of $\mathrm{V_{1-x}Ti_xAl_3}$ do not show any structural transition after a long time annealing at low temperature and our XRD measurements verified they are isostructural solid solutions.
The orientations of the single crystals were determined by Laue diffraction in a Photonic Science PSL-laue diffractometer. The compositions of V and Ti for the series were determined by Energy Dispersion Spectroscopy (EDS) in an X-Max 80. For the whole series, the ratio of Ti and V in the crystals is same as the starting elements of $1-x:x$ within $\pm2\%$ uncertainty, which is close to the estimated tolerance of the instrument ($\pm1\%$). Since our EDS and XRD measurements have verified the homogeneity of the V$_{1-x}$Ti$_x$Al$_3$ solid solutions \cite{raghavan2012ti}, we used the starting $x$ as the nominal $x$ in this paper.

\subsection*{Electrical and magnetic measurements}
The resistance, Hall effect, and PHE were performed in a Quantum Design Physical Property Measurement System (PPMS), using the standard four-probe technique with the silver paste contacts cured at room temperature. Temperature dependent resistance measurement showed that the whole series are metallic (See SI Appendix, Fig. S2) with a room temperature resistivity $\rho_{300K}$ approximately equaling $100~\mu\Omega \mathrm{cm}$. In order to avoid the longitudinal resistivity contribution due to voltage probe misalignment, the Hall resistivity was measured by sweeping the field from $-9$ T to $9$ T at various temperatures and then symmetrized as $\rho_{yx}(H)=[\rho(+H)-\rho(-H)]/2$. The PHE was measured in a fixed magnetic field ($\mu_0H=\pm~5~ \mathrm{T}$) and the sample was rotated so that the magnetic field direction was kept in the plane of the current and Hall contacts. To remove the regular Hall contribution and zero-field background, we determined the planar Hall resistivity as $\rho_{yx}^\mathrm{PHE}(H)=[\rho_{yx}^\mathrm{PHE}(+H)+\rho_{yx}^\mathrm{PHE}(-H)]/2-\rho_{yx}^\mathrm{PHE}(H=0)$. The experiment setting and data analysis method of $\rho_{yx}^\mathrm{PHE}$ ensure that there is no contribution from regular Hall effect and anomalous Hall effect. The Seebeck coefficient from $300$ K to $100$ K (See SI Appendix, Fig. S5) was measured in a home-built thermoelectric measurement system which uses constantan as the reference. The temperature-dependent molar susceptibility and Field-dependent magnetization of $\mathrm{VAl_3}$ were measured in Superconducting Quantum Interference Device (MPMS SQUID VSM) and the relevant results see SI Appendix, Fig. S6 and S7.

\subsection*{Electronic Calculation}
The electronic structures of $\mathrm{TiAl_3}$ and $\mathrm{VAl_3}$ were calculated using CAESAR \cite{ren1998caesar} with semi-empirical extended-Hückel-tight-binding (EHTB) methods \cite{hoffmann1963extended}. The parameters for Ti are $4s$: $\zeta=1.075$, $Hii=–8.97eV$; $4p$: $\zeta=0.675$, $Hii=–5.44eV$, and $3d$: $Hii=–10.810eV$; $\zeta1=4.550$, $Coefficient1=0.4206$; $\zeta2=1.400$, $Coefficient1=0.7839$; V are $4s$: $\zeta=1.300$, $Hii =–8.81eV$; $4p$: $\zeta=1.300$, $Hii=-5.52eV$, and $3d$: $Hii=–11.000eV$, $\zeta=4.750$, $Coefficient1=0.4755$; $\zeta2=1.700$,  $Coefficient2=0.7052$; ; Al are $3s$: $\zeta=1.167$, $Hii=–12.30eV$; $4p$: $\zeta=1.167$, $Hii=–6.50 eV$. Partial DOS and COHP calculations were performed by the self-consistent, tight-binding, linear-muffin-tin-orbital (LMTO) method in the local density (LDA) and atomic sphere (ASA) approximations \cite{andersen1986illustration}. Interstitial spheres were introduced in order to achieve space filling. The ASA radii as well as the positions and radii of these empty spheres were calculated automatically, and the values so obtained were all reasonable. Reciprocal space integrations were carried out using the tetrahedron method. We computed electronic structures using the projector augmented wave method \cite{blochl1994projector,kresse1999ultrasoft} as implemented in the VASP \cite{kresse1996efficient} package within the generalized gradient approximation (GGA) schemes \cite{perdew1996generalized}. A $15\times15\times15$ MonkhorstPack k-point mesh was used in the computations with a cutoff energy of $500$ eV. The spin orbital coupling (SOC) effects were included in calculations self-consistently.

\subsection*{Data Availability}
All data is contained in the manuscript text and supporting information.

\begin{acknowledgments}
We are grateful to Xi-Tong Xu for his useful advice in experiment and Guang-Qiang Wang for his help in Laue orientation. This work was supported by the National Natural Science Foundation of China No. U1832214, No.11774007, the National Key R\&D Program of China (2018YFA0305601) and the Strategic Priority Research Program of Chinese Academy of Sciences (Grant No. XDB28000000). The work at LSU was supported by Beckman Young Investigator (BYI) program. T.-R.C. was supported by the Young Scholar Fellowship Program from the Ministry of Science and Technology (MOST) in Taiwan, under a MOST grant for the Columbus Program MOST109-2636- M-006-002, National Cheng Kung University, Taiwan, and National Center for Theoretical Sciences, Taiwan. This work was supported partially by the MOST, Taiwan, Grant MOST107-2627-E-006-001.This research was supported in part by Higher Education Sprout Project, Ministry of Education to the Headquarters of University Advancement at National Cheng Kung University (NCKU).

\end{acknowledgments}

\end{document}